\documentclass[conference]{IEEEtran}
\usepackage[T1]{fontenc}
\usepackage[utf8]{inputenc}
\usepackage{amsmath,amssymb,amsfonts}
\usepackage{graphicx}
\usepackage{url}
\usepackage{cite}
\usepackage{balance}
\usepackage{array}
\usepackage{hyperref}
\hypersetup{hidelinks}
\title{Clearing-Instant Dependence and Persistent Post-Fault Voltage Recovery in a Grid-Following Inverter: An EMT Case Study}
\author{\IEEEauthorblockN{Dileep Cheerla, Amirthagunaraj Yogarathnam, Divya Jain, and Satish Natti}
\IEEEauthorblockA{GridAxon, Inc., Lewisville, Texas, USA\\
\{dileep, araj, divya, satish\}@gridaxon.io}}
\begin{document}
\maketitle
\noindent\begin{minipage}{\columnwidth}
\footnotesize\emph{This work has been submitted to the IEEE for possible publication. Copyright may be transferred without notice, after which this version may no longer be accessible.}
\end{minipage}\par\vspace{1.0em}
\begin{abstract}
A short qualifying post-fault voltage interval does not necessarily establish persistent recovery in a grid-following (GFL) inverter. This EMT case study evaluates twenty-five distinct balanced-fault cases of an aggregated battery-energy-storage inverter operated in discharge. The primary twelve-case matrix crosses short-circuit ratio (SCR) 2/5, P/Q current priority, and clearing duration 100/200/300 ms. 

Seven additional SCR-2 P-priority clearing instants, two SCR-2 Q-priority instants, and three additional SCR-5 P-priority severe-fault instants supplement the crossed matrix; one SCR-5 Q-priority severe-fault test completes the control. A 20-ms first in-band interval is distinguished from lasting recovery through the 10-s simulation horizon. 

At SCR 2, P-priority lasting \(\pm\)2\% voltage-recovery times range from 84.6 to 1216.6 ms; a 5-ms change in fault duration from 210 to 215 ms raises the measured lasting recovery from 413.8 to 995.3 ms. For the SCR-5 severe-fault controls (|Zf | = 0.050 pu), lasting P-priority recovery is 23.4 and 23.0 ms for 200- and 220-ms faults, compared with 510.3 and 990.3 ms at SCR 2; the 300-ms SCR-5 P/Q cases recover within 36--42 ms. This challenges a fault-depth-only explanation for the observed delay, without isolating grid strength or an internal synchronization mechanism. The findings motivate recovery-persistence reporting and careful treatment of clearing instant and fault severity in EMT case studies. \end{abstract}
\section{Introduction}
 Grid-following (GFL) inverter-based resources inject controlled currents into power networks whose voltage and impedance characteristics can change sharply during faults. When converter current reaches its limit, active- and reactive-current priority lead to different current allocations, faulted-voltage responses, and active-power trajectories. Grid impedance and synchronization dynamics can further complicate the response \cite{ref2}--\cite{ref7}, \cite{ref9}. At the network-operation timescale, battery-integrated photovoltaic dispatch has been examined for coordinated scheduling and economic performance \cite{ref12}; that work does not address subsecond fault recovery. Quantitative EMT conclusions require a defined inverter model and operating condition. 

Analytical studies relate GFL fault-ride-through current injection and clearing time to transient-stability regions \cite{ref13}, and examine PLL-based fault-clearing-angle formulations \cite{ref10}. The present work instead evaluates measured recovery persistence without identifying internal stability regions or PLL angles. The criterion used to judge post-fault recovery also matters. A positive-sequence voltage may enter its prefault tolerance band for a short interval and subsequently leave it. Similarly, fast active-power restoration need not demonstrate sustained voltage restoration. In a time-domain model, these distinctions matter during and after clearing. Control-layer limits also affect reported reactive-power recovery in EMT-- RMS benchmarking \cite{ref11}, without addressing clearing-instant sensitivity. 

Previous EMT work has examined the effects of SCR, network X/R, and current priority at a single 150-ms fault duration \cite{ref1}. The present study addresses a distinct question: how do fault-clearing duration and grid condition affect the difference between first-qualifying and lasting post-fault recovery? The core twelve-case EMT matrix crosses two short-circuit ratios (SCR 2 and 5), three durations (100, 200, and 300 ms), and two current-priority modes. Thirteen distinct targeted cases probe a 5-ms transition interval and a second physical fault impedance. The application is an aggregated BESS model in discharge, used here as a representative GFL inverter case rather than as evidence about all BESS or PV implementations. 

The contributions are (i) explicit comparison of first qualifying and finite-horizon lasting voltage recovery, (ii) localization of a pronounced recovery-time change to a 5-ms clearing interval from ten SCR-2 P-priority durations, and (iii) SCR-5 reduced-impedance control runs at three clearing instants, alongside voltage-tolerance and waveform-based interpretation. The work does not claim a generic stability boundary, IEEE Std 2800 compliance, or identification of an individual internal controller state. 
\section{Model and Test Conditions}
 
\subsection{Study system and operating points}
\begin{table}[!t]
\caption{EMT conditions (12 primary and 13 distinct targeted tests)}\label{tab:conditions}
\centering\scriptsize\setlength{\tabcolsep}{3pt}
\begin{tabular}{p{0.43\columnwidth}p{0.48\columnwidth}}\hline
\textbf{Quantity}&\textbf{Setting}\\\hline
Grid SCR; X/R & 2 and 5; 8\\
Operating power command & $P_{\mathrm{cmd}}=+0.84$ pu (discharge)\\
Reactive reference, SCR 2 & 0.108 pu\\
Reactive reference, SCR 5 & 0.097 pu\\
Fault type and location & Balanced three-phase, POI\\
Primary fault impedance; X/R & $|Z_f|=0.125$ pu; approx. 8\\
Fault application & 5.00 s\\
Core clearing & 5.10, 5.20, or 5.30 s\\
LVRT priority modes & P-priority; Q-priority\\
LVRT activation / hysteresis & Fixed identically across cases\\
LVRT reactive-current gain & Fixed identically across cases\\
Converter command-current limit & 1.0 pu\\
Simulation step / recorded interval & 20 $\mu$s / 100 $\mu$s (25 cases)\\
Additional SCR-2/P durations & 180, 205, 210, 215, 220, 280, 320 ms\\
Additional SCR-2/Q durations & 210, 220 ms\\
Additional SCR-5 fault & $|Z_f|=0.050$ pu; 200/220 ms P, 300 ms P/Q\\
Run horizon & 10.0 s\\\hline
\end{tabular}
\end{table}

 A PSCAD electromagnetic-transient (EMT) representation of an aggregated GFL BESS comprises an inverter-level P/Q controller, synchronization and LVRT logic, current limiting, equivalent LV/MV and collection-network impedances, a main transformer, and a Thevenin grid connected at the point of interconnection (POI). A related fixed-duration impedance-sensitivity study used a similar EMT test architecture \cite{ref1}. The twenty-five distinct cases reported here evaluate clearing-time

and recovery-persistence questions distinct from the fixed-duration analysis in that study. Active- and reactive-power references are applied directly; no separate external plant-level power-controller dynamics are evaluated. The implementation is presented by functional behavior without identifying a commercial product or disclosing internal proprietary control diagrams. 

Grid strength and impedance angle are represented on the nominal POI-voltage and aggregate power bases as

\begin{align}
\mathrm{SCR}&=\frac{S_{\rm sc}}{S_n},\quad |Z_{\rm th,pu}|=\frac{1}{\mathrm{SCR}}, \label{eq:scr}\\
R_{\rm th,pu}&=\frac{|Z_{\rm th,pu}|}{\sqrt{1+k^2}},\quad X_{\rm th,pu}=kR_{\rm th,pu},\quad k=\frac{X}{R}.\label{eq:grid}
\end{align}

Here \(S_{\rm sc}\) is the grid three-phase short-circuit MVA and \(S_n\) is the modeled aggregate rating. Equations (1)--(2) describe the external Thevenin grid, not proprietary inverter controls. 

On these bases, a physical phase-rms voltage is converted by \(V_{1,\mathrm{pu}}=|V_1|/(V_{LL,b}/\sqrt{3})\) and three-phase power by \((P_1+jQ_1)_{\mathrm{pu}}=3V_1I_1^*/S_n\). The same bases are used across the 25 runs. 

The twelve primary cases are arranged at SCR 2 and 5 with X/R = 8, Pcmd = +0.84 pu, and case-specific Qcmd = 0.108 pu (SCR 2) or 0.097 pu (SCR 5) to hold pre-fault POI reactive exchange near zero. Over 4.80--4.95 s, the reconstructed POI pre-fault values are approximately (P1,pre, Q1,pre, V1,pre) = (0.8387, 0.0003, 0.9563) pu at SCR 2 and (0.8394, \(-\)0.0005, 1.0084) pu at SCR 5. The analysis uses Sb = 238.1 MVA and VLL,b = 138.0 kV (Zb \(\simeq\) 80.0 \(\Omega\)), cross-checked against raw prefault power and phase-voltage channels and \cite{ref1}. These are independently reconstructed analysis bases; the original project parameter record is not available for certification. All fault impedances below use this base. 

Only BESS discharge is studied. Relative POI voltage is normalized by each grid condition's own prefault value for the cross-SCR recovery comparisons. 
\subsection{LVRT modes and disturbance}
 The tested P-priority and Q-priority LVRT settings form active- and reactive-current requests from operating-point references and an undervoltage-dependent reactive-current increment. Under a 1.0-pu commanded-current ceiling, P-priority preserves the requested d-axis current to the extent allowed by the limit, whereas Q-priority preserves the requested qaxis current. For either mode, the generic commanded-current circle gives

\begin{equation}\left(i_d^*\right)^2+\left(i_q^*\right)^2\le I_{\max}^2,\quad I_{\max}=1.0\;\mathrm{pu}.\label{eq:circle}\end{equation}

In an idealized priority allocation, once one component \(i^*_{\mathrm{pri}}\) is selected, the other is bounded by

\begin{equation}\left|i_{\rm rem}^*\right|\le \sqrt{I_{\max}^2-\left(i_{\rm pri}^*\right)^2}.\label{eq:remainder}\end{equation}

This is a geometric current-capability interpretation rather than an asserted implementation-specific limiter algorithm. Both priority modes use the same fixed LVRT request and release settings. These descriptions concern the functional priority ordering, not an independently validated disclosure of internal controller algorithms; no control parameters other than the priority selection are intentionally changed within each paired comparison. 

A common timed command applies a balanced three-phase POI fault at 5.00 s and commands removal at the stated clearing time. The exported event channels do not identify whether the simulated three-phase fault switch interrupts simultaneously or independently at phase-current zeros; breaker-contact traces were not recorded. The fault X/R \(\simeq\) 8 is held to represent an inductive fault branch and to maintain a consistent reference angle with the tested grid; point-on-wave and interruption-method sensitivity are not separately tested. 

A common three-phase fault of |Zf | = 0.125 pu and fault X/R \(\simeq\) 8 is used in the twelve primary cases and the nine targeted SCR-2 cases. Four additional SCR-5 tests instead use |Zf | = 0.050 pu with the same fault X/R. Within each fixed SCR, fault-impedance, and priority subgroup, clearing time is the only test variable. Across SCR groups, the Thevenin impedance changes while the physical fault stays fixed, so passive retained voltage also changes. For an unloaded, linear Thevenin divider, the passive fault-voltage ratio is  

\begin{equation}\left|\frac{V_{\rm ret,pass}}{V_{\rm th}}\right|=\left|\frac{Z_f}{Z_{\rm th}+Z_f}\right|.\label{eq:divider}\end{equation}

Using the tested impedance magnitudes and angles, the passive ratio is approximately 0.20 at SCR 2 versus 0.385 at SCR 5. This network-only check does not predict the actively controlled inverter's actual faulted POI voltage. Therefore the two-grid comparison cannot distinguish pure SCR effects from coincident changes in fault severity. Responses at clearing may also depend on the time-varying fault-period controller state. 
\section{Measurements and Recovery Criteria}
 
\subsection{Positive-sequence reconstruction}
 Each EMT case is solved at 20 \(\mu\)s. All twenty-five distinct cases record three-phase POI voltages and currents every 100

\(\mu\)s. A nominal one-cycle, 60-Hz rectangular discrete Fourier transform (167 samples for 100-\(\mu\)s recordings) reconstructs phase RMS phasors, with the positive-sequence components V1 and I1 used to calculate

\begin{equation}S_{1,\rm pu}=P_{1,\rm pu}+jQ_{1,\rm pu}=\frac{3V_1 I_1^*}{S_n}.\label{eq:power}\end{equation}

Here V1 and I1 are physical phase RMS positive-sequence phasors, \(S_n\) is the aggregate three-phase MVA base, and positive P1,pu denotes injection toward the grid. Positive Q1 denotes reactive injection at the POI; a negative inverter-frame \(i_q^*\) may denote positive injection under the model's separate dq sign convention. The per-unit powers are written as P1 and Q1 below. A centered window is used to describe fault-period traces away from event boundaries, while recovery measurements are timestamped at the trailing window end, using the window endpoint. The trailing timestamp is 8.3 ms later than the centered timestamp for the 167-sample recordings. The reported recovery values use trailing timestamps; the tables consistently use trailing-end timestamps. A fixed-frequency DFT can have leakage under off-nominal converter frequency, particularly during the longer P-priority faults; fault-period phasor magnitudes should therefore be read as approximate measurements. 

A trailing average of instantaneous positive-sequence voltage magnitude reproduces recovery times for the five new P-priority runs within 0.1 ms, although estimated peak overvoltage is estimator-dependent. All reported cases use a 100-\(\mu\)s sampling interval; no sub-millisecond differences in recovery are interpreted as separate physical effects. 
\subsection{First and lasting recovery}

Let tc denote fault clearing, Te = 10 s the end of the record, V1,pre the pre-fault positive-sequence POI voltage, and Tq = 20 ms the qualification interval. Define the in-band indicator

\begin{equation}b_\epsilon(t)=\mathbf{1}\!\left(\left|\frac{V_1(t)}{V_{1,\rm pre}}-1\right|\le\epsilon\right),\quad\epsilon=0.02.\label{eq:indicator}\end{equation}

where 1(\(\cdot\)) is one if the condition holds and zero otherwise. Over the sampled post-clearing trajectory, the recovery delays are

\begin{align}
t_{V,\mathrm{first}}&=\min\left\{t-t_c:\; b_\epsilon(s)=1,\;\forall s\in[t,t+T_q]\right\},\label{eq:first}\\
t_{V,\mathrm{last}}&=\min\left\{t-t_c:\; b_\epsilon(s)=1,\;\forall s\in[t,T_e]\right\}.\label{eq:last}
\end{align}

For \(t_{V,\mathrm{first}}\) the full qualification interval must fit within the recorded horizon; an undefined metric is reported if no qualifying interval exists. For \(t_{V,\mathrm{last}}\) the requirement is continuous satisfaction at all recorded samples from that time through Te. A late out-of-band sample can shift \(t_{V,\mathrm{last}}\) by many milliseconds, so the quantity is tolerance- and horizondependent. The gap \(\Delta\)tV,persistence = \(t_{V,\mathrm{last}}\) \(-\)\(t_{V,\mathrm{first}}\) measures elapsed time between the two milestones, not cumulative out-of-band duration. Lasting recovery here does not demonstrate asymptotic convergence beyond the finite record.

\begin{table}[!t]
\caption{Primary twelve-case recovery after clearing (ms)}\label{tab:primary}
\centering\scriptsize\setlength{\tabcolsep}{3.5pt}
\begin{tabular}{cccrrr}\hline
SCR & Duration & Priority & $t_{P95,\mathrm{first}}$ & $t_{V,\mathrm{first}}$ & $t_{V,\mathrm{last}}$\\
& (ms) & & (ms) & (ms) & (ms)\\\hline
2 & 100 & P & 43.9 & 84.6 & 84.6\\
2 & 100 & Q & 55.0 & 55.2 & 55.2\\
2 & 200 & P & 30.1 & 35.9 & 510.3\\
2 & 200 & Q & 54.8 & 55.0 & 55.0\\
2 & 300 & P & 29.9 & 851.0 & 910.8\\
2 & 300 & Q & 55.1 & 68.2 & 68.2\\
5 & 100 & P & 20.1 & 22.3 & 22.3\\
5 & 100 & Q & 21.8 & 21.0 & 21.0\\
5 & 200 & P & 20.0 & 22.3 & 22.3\\
5 & 200 & Q & 21.9 & 23.7 & 23.7\\
5 & 300 & P & 20.1 & 22.3 & 22.3\\
5 & 300 & Q & 21.9 & 21.0 & 21.0\\\hline
\end{tabular}
\end{table}

Active-power restoration is characterized by \(t_{P95,\mathrm{first}}\), the first 20-ms interval with P1(t) \(\ge\) 0.95P1,pre. Time is always measured from the respective clearing event, rather than from the common 5.00-s application time. Neither the 20-ms metric nor the lasting metric is asserted to be an IEEE Std 2800 compliance test. 
\section{Results}
 
\subsection{Recovery differences across grid conditions}
\begin{figure*}[!t]
\centering
\includegraphics[width=\textwidth]{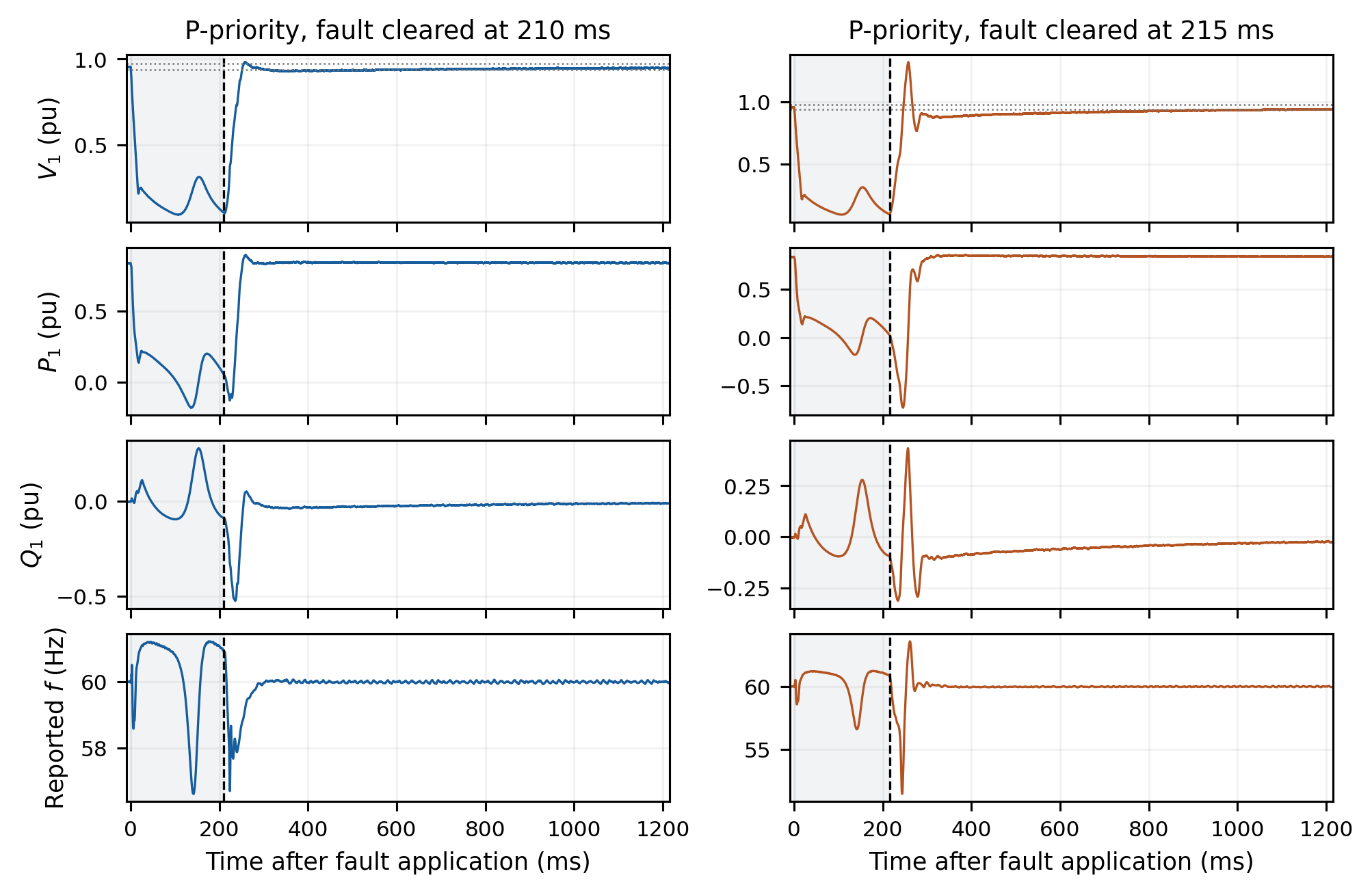}
\caption{SCR-2 P-priority measurements for 210- and 215-ms clearing, aligned to fault application. Top to bottom: V1, P1, Q1 at the POI, and reported controller frequency. Shading denotes the applied fault, dashed lines mark clearing, and dotted horizontal lines on the voltage plots denote the prefault-relative \(\pm\)2\% band. The 215-ms clearing is followed by an approximately \(-\)0.6-pu active-power excursion, a reported-frequency dip near 52 Hz, and a positive-sequence voltage crest near 1.3 pu; the 210-ms case lacks this combination.}\label{fig:1}
\end{figure*}

 For every late-fault voltage comparison, define \(\overline{V}_{1,f}\) as the mean positive-sequence magnitude on [tc \(-\)45 ms, tc \(-\)5 ms), where tc is the corresponding clearing instant. The 5-ms exclusion reduces mixing with the clearing discontinuity. A common definition is essential because P-priority oscillations make fixed-clock and clearing-relative averages noninterchangeable. Table II summarizes the twelve primary cases. The supplementary thirteen-case tests are reported separately to preserve the distinction between the crossed matrix and targeted controls. At SCR 2, P-priority lasting \(\pm\)2\% POI voltage recovery is 84.6, 510.3, and 910.8 ms for 100-, 200-, and 300-ms faults, respectively; corresponding Q-priority results are 55.2, 55.0, and 68.2 ms. At SCR 5, P-priority recovery is 22.3 ms for all three durations, while Q-priority is approximately 21--24 ms. Thus the marked increase in recovery duration under weak-grid P-priority is not reproduced in the stronger-grid fixed-fault conditions. The sampled outcomes are listed in Table II. 

Within the SCR-2 subgroup, the 200-ms P-priority case is especially instructive. Its first valid 20-ms voltage interval starts at 35.9 ms, yet the voltage later departs from the selected band and only remains inside it after approximately 510.3 ms. The 300-ms case first qualifies at 851.0 ms but has another subsequent out-of-band excursion, giving \(t_{V,\mathrm{last}}\) = 910.8 ms. This shows why an early-window recovery criterion alone may understate the observed recovery tail for this model. The Q-priority cases do not exhibit this first-versus-lasting discrepancy at the \(\pm\)2\% threshold. For the 300-ms tests, maximum one-second post-clearing positive-sequence rms voltage (normalized to nominal phase rms voltage) is approximately 0.945/1.155 pu for SCR-2 baseline P/Q, 1.013/1.026 pu for SCR-5 baseline P/Q, and 1.031/1.055 pu for SCR-5 reduced-fault P/Q. The maximum sampled instantaneous phase voltage, normalized instead to the nominal phase peak, is respectively 1.349/1.456, 1.087/1.121, and 1.218/1.160 pu. These are different estimands: the phase-peak measure includes subcycle excursions and must not be labeled a positive-sequence rms overvoltage. 
\subsection{Why the recovery definitions change the interpretation}

The distinction between first-window and lasting recovery is not merely terminological. The first-window test identifies one 20-ms sequence of acceptable voltage samples; the lasting test evaluates every subsequent observation through 10 s. Neither definition alone establishes small-signal stability, yet together they expose a nonpersistent early return that would otherwise be obscured by a single recovery number. For SCR 2 at 200 ms under P-priority, the interval between the two reported recovery times is 510.3\(-\)35.9 = 474.4 ms. The corresponding gap at 300 ms is 910.8 \(-\) 851.0 = 59.8 ms. Thus the 200-ms case exhibits a much larger first-to-lasting discrepancy even though its final recovery is earlier than in the 300-ms case. These are different characteristics: the elapsed time to the first qualifying window and the time of the last observed departure need not vary in parallel. 

The reported gap does not imply that voltage is continuously outside its band; the 10-s observation horizon and selected tolerance define its interpretation. 
\subsection{Active-power recovery is not lasting voltage recovery}
 Despite the long voltage tail, P-priority reaches the first active-power recovery threshold in 43.9, 30.1, and 29.9 ms, respectively. In the 300-ms case, the separation between \(t_{P95,\mathrm{first}}\) and \(t_{V,\mathrm{last}}\) is about 881 ms. Thus, the first qualifying active-power interval occurs much earlier than lasting voltage recovery in this case. The priority logic tends to retain active-current request under saturation, but the observed timing difference alone does not identify which internal dynamic state controls the later voltage tail. 

At SCR 2, Q-priority \(t_{P95,\mathrm{first}}\) stays near 55 ms across all three durations. Q-priority also yields higher late-fault positive-sequence voltage than P-priority in every pair, although the magnitude of the P-priority late-fault average changes with fault duration because its fault response is nonstationary. For SCR 2, the late-fault V1 values for the 100-, 200-, and 300-ms P-priority cases are approximately 0.131, 0.225, and 0.250 pu, respectively, using the same 40-ms clearing-relative window, [tc \(-\) 45 ms, tc \(-\) 5 ms). The paired Q-priority values are approximately 0.298--0.299 pu. These changing window averages should not be taken as a simple steady-state voltage improvement with longer faults. 
\subsection{Stronger-grid control comparison}
 For SCR 5, the prefault V1 is about 1.0084 pu, compared with 0.9563 pu at SCR 2. The late-fault 300-ms averages are about 0.398 pu for P-priority and 0.444 pu for Q-priority, versus 0.250 pu and 0.299 pu, respectively, at SCR 2. The stronger-grid P-priority response retains positive fault-period active power (approximately 0.394 pu during the late 300-ms window) and shows no analogous long post-fault voltage tail. The respective POI voltages are normalized to their own prefault magnitudes in the recovery analysis; no common pre-fault voltage is imposed. The 300-ms P-priority lasting-recovery difference is approximately 888.5 ms between the two grid conditions, but it is a fixed-physical-fault contrast, not a constant-retained-voltage SCR sensitivity. The stronger grid reduces Thevenin impedance and changes retained voltage during the fault, so their separate contributions cannot be resolved by the original twelve-case matrix. 
\subsection{Measured current angle and fault-period dynamics}
\begin{figure*}[!t]
\centering
\includegraphics[width=\textwidth]{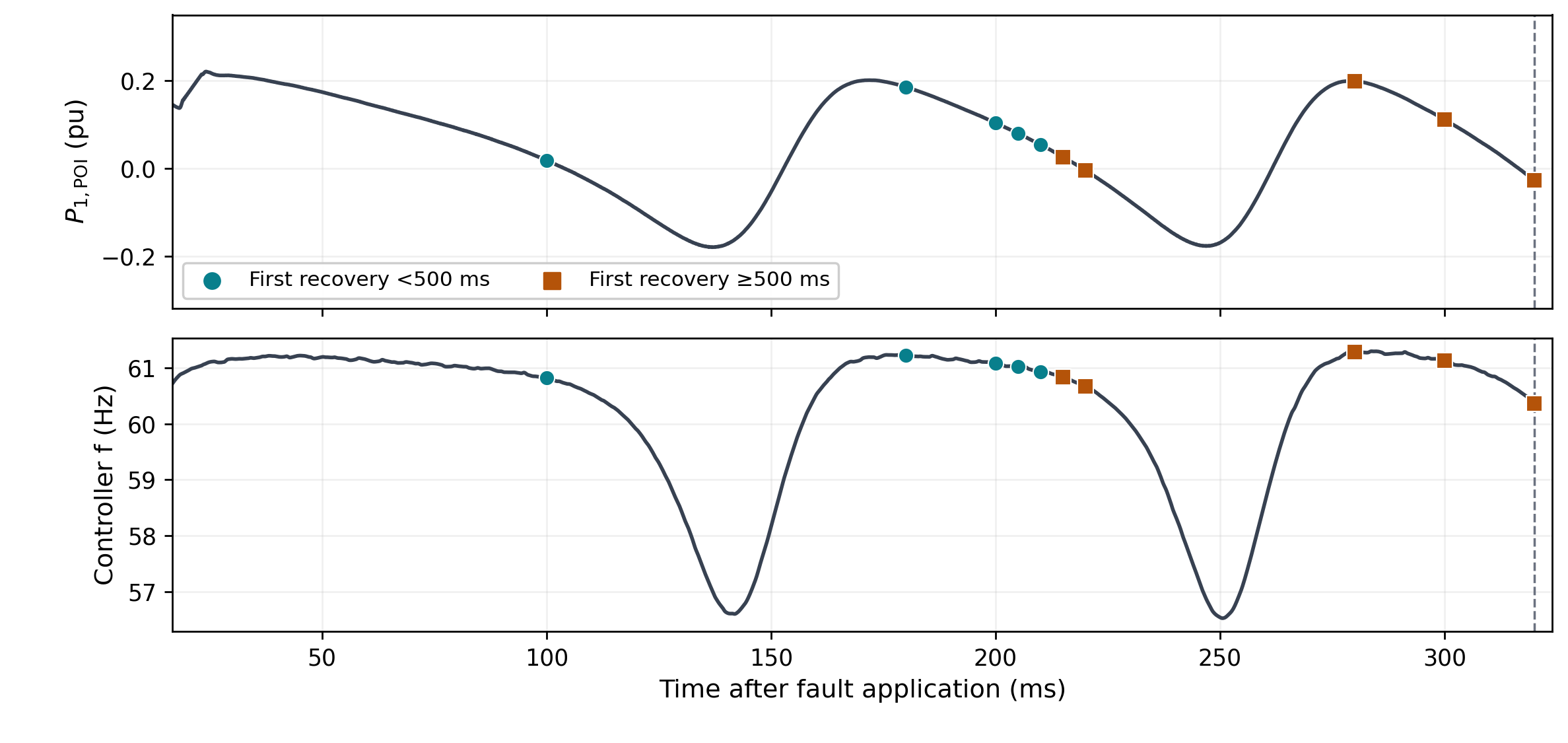}
\caption{SCR-2 P-priority POI positive-sequence power (trailing one-cycle DFT) and reported controller frequency during the 320-ms fault. The trace ends before the programmed clearing; no post-clearing samples are shown. Markers denote ten tested clearing instants (the 320-ms marker uses the last pre-clearing sample), grouped by the separate-run first qualifying voltage-recovery outcome. The earliest displayed power is after the first complete fault-period DFT window.}\label{fig:2}
\end{figure*}

 Figure 2 places ten clearing instants on a recorded 320-ms fault-period trajectory, truncated before the commanded 320-ms clearing. Its trailing-window positive-sequence power is plotted only after one complete post-inception cycle; thus no point includes post-clearing waveform samples. During the 300-ms P-priority case the reconstructed POI active power crosses into the absorption quadrant, reaching approximately \(-\)0.18 pu. The reported inverter/controller-frequency signal also exhibits a sizable excursion (about 56.5--61.3 Hz over the defined fault-evaluation portion), compared with approximately 58.7--60.0 Hz for Q-priority. As an independent electrical consistency check, the measured positive-sequence POI current-to-voltage angle is \(\arg(I_1V_1^*)\) \(\approx\) \(-\)118.4\(^\circ\) at 5.255 s in the 300-ms P-priority case, when P1 \(\approx\) \(-\)0.121 pu. The simultaneously recorded inverter-frame command angle is \(\operatorname{atan2}(i_q^*,i_d^*)\) \(\approx\) \(-\)31.4\(^\circ\) . For a frame-consistency check, define \(\delta\) = wrap180[\(\arg(I_1V_1^*)\)\(-\)\(\operatorname{atan2}(i_q^*,i_d^*)\)]. The median prefault value over 4.80--4.95 s is +7.3\(^\circ\) (median absolute deviation 0.06\(^\circ\) ). At 5.255 s, \(\delta\) \(\approx\) \(-\)87.1\(^\circ\) ; subtracting the prefault offset gives a change of approximately \(-\)94.4\(^\circ\) . The approximately \(-\)94.4\(^\circ\) prefault-referenced change is consistent with an approximately quarter-cycle rotation of the measured current orientation relative to the commanded frame, but does not identify the underlying mechanism. This is a relative phasor-versus-command-angle diagnostic, not a measured PLL-to-grid angle: current tracking, voltage depression, electrical location, and dq conventions can all contribute dynamically. The recorded frequency excursion is descriptive only and is not integrated to infer a bounded synchronization angle. The 100-ms fault clears before the later P-priority power reversals develop. In the 300-ms fault the recorded late-fault command vectors are approximately (\(i_d^*\), \(i_q^*\)) = (0.854, \(-\)0.521) pu for P-priority and (0, \(-\)1.000) pu for Q-priority, consistent with the documented 1.0-pu current ceiling. These are recorded current commands and a reported controller-frequency signal, not independent identification of internal PLL-angle dynamics; no loss-of-synchronism threshold is inferred. 

At approximately 200 ms after clearing, the longer P-priority cases have reactive exchange of about \(-\)0.033 pu (200-ms fault) and \(-\)0.069 pu (300-ms fault), with corresponding POI voltages near 0.931 and 0.903 pu. The matched Q-priority cases show reactive exchange near \(-\)0.003 pu and voltage near 0.954 pu. The 300-ms pair is a useful illustrative case. A weak, predominantly inductive network provides a physically plausible voltage sensitivity to reactive-power exchange. In a local power-flow approximation, network voltage changes reflect both active- and reactive-power perturbations, with the latter receiving greater weight when X/R is large. However, the observed negative Q1 and depressed V1 can both be consequences of the evolving inverter/network dynamics; they do not prove that reactive-power absorption caused the recovery delay. Confirming causality would require targeted controller-state measurements or controlled perturbations. 
\subsection{Targeted clearing-instant and fault-severity controls}
\begin{table}[!t]
\caption{Thirteen targeted tests: clearing-referenced recovery (ms)}\label{tab:targeted}
\centering\scriptsize\setlength{\tabcolsep}{3pt}
\begin{tabular}{ccrrrrr}\hline
SCR & Mode & $T_f$ & $|Z_f|$ & $t_{P95}$ & $t_{V,\mathrm{first}}$ & $t_{V,\mathrm{last}}$\\
 & & (ms) & (pu) & (ms) & (ms) & (ms)\\\hline
2 & P & 180 & .125 & 29.2 & 37.5 & 530.1\\
2 & P & 205 & .125 & 29.9 & 30.2 & 505.1\\
2 & P & 210 & .125 & 40.8 & 51.4 & 413.8\\
2 & P & 215 & .125 & 77.0 & 948.1 & 995.3\\
2 & P & 220 & .125 & 59.2 & 964.3 & 990.3\\
2 & P & 280 & .125 & 39.1 & 871.9 & 930.3\\
2 & P & 320 & .125 & 58.5 & 1216.6 & 1216.6\\
2 & Q & 210 & .125 & 54.7 & 71.7 & 71.7\\
2 & Q & 220 & .125 & 54.0 & 54.8 & 54.8\\
5 & P & 200 & .050 & 23.8 & 23.4 & 23.4\\
5 & P & 220 & .050 & 23.5 & 23.0 & 23.0\\
5 & P & 300 & .050 & 27.8 & 42.3 & 42.3\\
5 & Q & 300 & .050 & 38.4 & 36.0 & 36.0\\\hline
\end{tabular}
\end{table}
\begin{figure}[!t]
\centering
\includegraphics[width=\columnwidth]{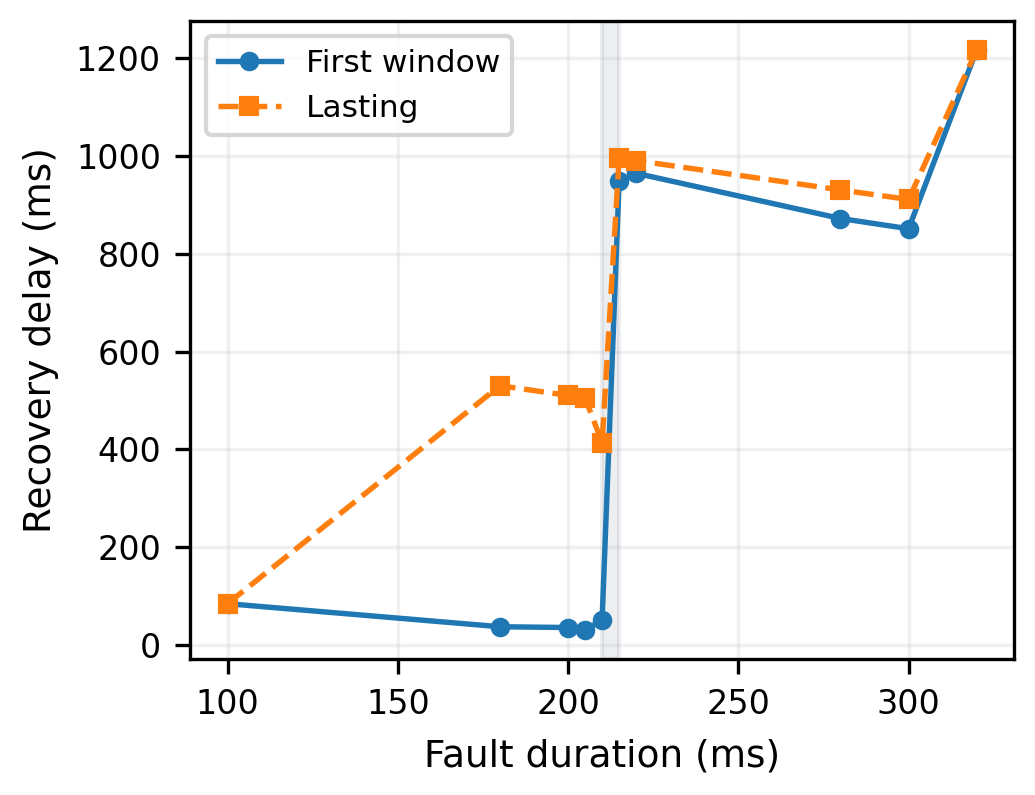}
\caption{SCR-2 P-priority first and lasting recovery for ten discrete clearing instants, including the 210--215-ms interval; joining lines do not establish a continuous boundary.}\label{fig:3}
\end{figure}

 \emph{The 210--215-ms divergence.}\  In the 215-ms P-priority case, the post-clearing excursion in P1 reaches approximately \(-\)0.6 pu, the reported controller-frequency signal falls near 52 Hz, and the positive-sequence voltage crest approaches 1.3 pu; the 210-ms case does not show this combination (Fig. 1). The sampled phase-voltage maximum within 1 s after 215-ms clearing is about 1.75 pu on the nominal phase-crest base at 28.3 ms after commanded clearing, whereas the positive-sequence rms peak is about 1.32 pu at 42.5 ms. These are distinct measurements and neither establishes a hardware overvoltage. Because the fault-contact interruption method has not been independently verified, the post-clearing transient as a whole, not only its sampled voltage peak, may partly depend on how the modeled fault branch is interrupted. Seven additional SCR-2 P-priority runs probe 180, 205, 210, 215, 220, 280, and 320 ms. Together with the primary 100/200/300-ms cases, ten clearing durations are available for P-priority. The first qualifying and lasting recovery delays at 210 ms are 51.4 and 413.8 ms, but at 215 ms they become 948.1 and 995.3 ms, respectively. Thus a pronounced difference occurs between these sampled 5-ms-separated instants; it is not a verified continuous threshold or critical clearing angle. The fault-period active-power signal exhibits repeated swings separated by roughly 0.11 s, but the recovery pattern does not repeat across successive swings: clearings 180--210 ms have substantially earlier first-window recovery than clearings 280--320 ms at broadly comparable positions after successive active-power troughs. Therefore, oscillation phase alone is insufficient; accumulated or otherwise evolving internal state is a plausible contributor, although it is unobserved and cannot be identified from these channels. Two additional SCR-2 Q-priority cases at 210/220 ms exhibit no analogous long recovery tail. Their 100-\(\mu\)s outputs yield lasting voltage recovery of 71.7 and 54.8 ms, respectively, with no corresponding extended recovery tail. 

For the matched-passive-severity control, the SCR-5 fault magnitude is reduced from 0.125 to 0.050 pu while SCR and X/R remain fixed. The recorded equivalent-fault branch signals verify the reduced fault magnitude and nominal fault X/R \(\simeq\) 8. The nominal passive divider is thus approximately matched to the SCR-2 baseline case, but the measured late-fault voltages are not equal: at 300 ms, SCR-2 baseline P/Q voltages are approximately 0.250/0.299 pu, whereas SCR-5 reduced-fault P/Q voltages are 0.165/0.240 pu, using the same 40-ms interval ending 5 ms before clearing for every 300-ms case. Nevertheless the SCR-5 reduced-fault lasting recovery values at 300 ms remain 42.3/36.0 ms for P/Q versus 910.8/68.2 ms at SCR 2. New SCR-5 reduced-fault P-priority tests at 200 and 220 ms recover in 23.4 and 23.0 ms, respectively, versus 510.3 and 990.3 ms at SCR 2. Equal passive-divider ratios do not guarantee identical realized voltage sags. This control challenges a retained-voltage-depth-only explanation but does not isolate SCR from differences in physical fault impedance, prefault voltage, phase at clearing, or the controller response. In particular, a single severe-fault current-suppression pulse can occur under both SCRs, while their long-term recovery outcomes differ. 
\subsection{Voltage-tolerance sensitivity}
\begin{figure}[!t]
\centering
\includegraphics[width=\columnwidth]{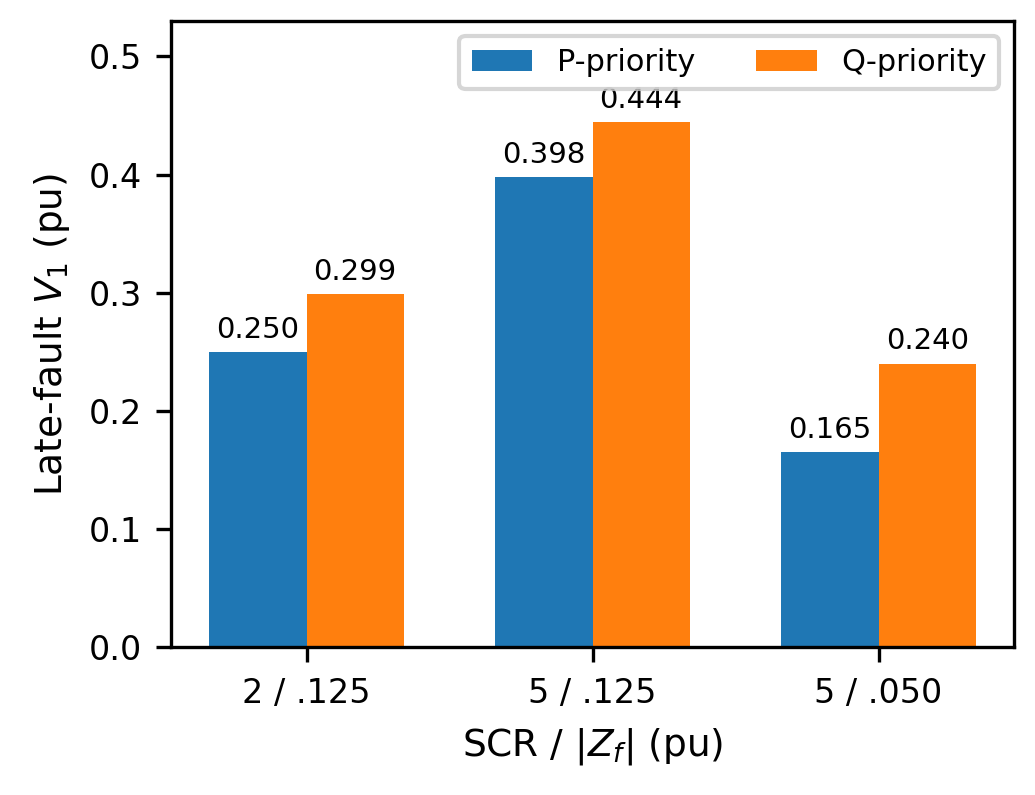}
\caption{Measured late-fault voltage for 300-ms tests, computed uniformly as mean positive-sequence POI voltage over [(tc \(-\) 45) ms,(tc \(-\) 5) ms). Different realized faulted operating points remain after nominal passive-divider matching.}\label{fig:4}
\end{figure}
\begin{table}[!t]
\caption{SCR-2 lasting POI voltage recovery versus tolerance (ms)}\label{tab:tolerance}
\centering\scriptsize\setlength{\tabcolsep}{5pt}
\begin{tabular}{ccrrr}\hline
Duration (ms) & Priority & $\pm1\%$ & $\pm3\%$ & $\pm5\%$\\\hline
100 & P & 104.6 & 63.2 & 57.0\\
100 & Q & 96.1 & 53.8 & 39.9\\
200 & P & 997.3 & 151.0 & 29.6\\
200 & Q & 96.8 & 53.6 & 39.7\\
300 & P & 1495.9 & 623.1 & 278.7\\
300 & Q & 96.8 & 54.0 & 39.9\\\hline
\end{tabular}
\end{table}

A fixed \(\pm\)2\% band is useful for comparison but does not uniquely define recovery. Table IV repeats lasting recovery using \(\pm\)1\%, \(\pm\)3\%, and \(\pm\)5\% of the common pre-fault voltage. The 300-ms P-priority tail remains markedly longer than Q-priority at every listed tolerance, so that particular contrast does not depend on the headline \(\pm\)2\% choice. The 200-ms P-priority result is more tolerance-sensitive: it is 997.3 ms under \(\pm\)1\%, 151.0 ms under \(\pm\)3\%, and only 29.6 ms under \(\pm\)5\%. Therefore the headline 510.3-ms value describes a specified narrow voltage-band persistence metric, not a general claim of half-second voltage collapse. 
\subsection{Event-flag and end-of-run checks}

The examined SCR-2 recordings contain a brief, approximately 5-ms assertion of the model output labeled disconnection near fault inception (about 5.0016--5.0066 s). The output is documented as a protection-related disconnection signal. The three-phase POI instantaneous-current vector has a mean magnitude over that 5-ms interval of approximately 0.15 times its prefault mean in the SCR-2 baseline 300-ms P-priority case; the corresponding SCR-5 baseline case is approximately 1.03, and the SCR-5 reduced-fault P-priority case approximately 0.12. These ratios are computed from \(\sqrt{(i_a^2+i_b^2+i_c^2)/3}\), not a one-cycle I1 estimator, whose filter cannot resolve a 5-ms pulse. The deep-sag cases therefore exhibit a brief momentary-cessation-like current-suppression interval, distinct from a sustained trip. The output asserts near the transient, but without the underlying protection-bit vector or exact terminal-current logic this is not certified as formal grid-code momentary cessation or attributed to a particular protection mechanism. No corresponding assertion is observed in the six SCR-5 cases with |Zf | = 0.125 pu; a similar short assertion appears in the four SCR-5 cases with |Zf | = 0.050 pu. The recorded LVRT flag asserts and subsequently deasserts in the cases examined; the short flag is also observed in the SCR-5 severe-fault cases despite rapid post-fault recovery. By the end of the 10-s simulation, the recorded POI power and voltage approach their pre-fault operating levels and controller frequency approaches 60 Hz; no persistent post-fault oscillation is apparent within this finite observation horizon. 
\section{Discussion and Limitations}
 The central finding is metric-dependent rather than a claim of newly discovered weak-grid power reversal: a first qualifying interval, an active-power restoration interval, and a lasting voltage-recovery interval can occur at very different times in the same response. The enlarged clearing-instant sweep provides new evidence that this separation is particularly pronounced for the tested weaker-grid, active-first configuration. Related fixed-duration work \cite{ref1} already documented weak-grid power reversal; the contribution here is the clearing-time and recovery-persistence comparison. 

The limitations are concentrated here. The 25 cases cover one discharge-mode inverter representation, two SCRs, two fault impedances at X/R = 8, and fixed controls; PV, charging, other controllers and field measurements are not tested. Nominal passive-sag matching does not isolate SCR or the state responsible for long recovery. Original model-base fields, the protection-bit source, PLL implementation, and breaker-contact interruption sequence cannot be independently certified from waveform exports; the analysis bases are nonetheless cross-checked numerically. Confidential controller gains, ramp limits, and internal diagrams are intentionally omitted, limiting independent reproduction. The command-time separation of the 215-ms peak from fault clearing does not exclude switch-model contributions or numerical artifacts, particularly without phase-contact traces and point-on-wave testing. The 5-ms current suppression is momentary-cessationlike, but no equipment/grid-code classification, IEEE 2800 compliance, or uninterrupted ride-through is asserted.

NERC EMT study guidance also emphasizes attention to voltage recovery and oscillatory behavior in interconnection evaluations \cite{ref7}, \cite{ref8}. For industrial power systems with inverter-based generation or storage, the principal engineering implication is a reporting discipline: first-window recovery and finite-horizon lasting recovery answer different questions, even when calculated from the same waveforms. For a strong fault-period transient, an early 20-ms qualification can be followed by later departure from the voltage band. Reporting tolerance, phasorwindow timestamp, qualification interval, and observation horizon makes recovery comparisons interpretable. A short power-recovery time must not be substituted for evidence of lasting voltage restoration. 
\section{Conclusion}
 Across twenty-five distinct EMT cases, the distinction between first qualifying and lasting post-fault POI-voltage recovery proved consequential for the tested GFL inverter in discharge. At SCR 2 under P-priority, discrete clearing times produced non-monotonic lasting-recovery delays from 84.6 to 1216.6 ms. SCR-5 reduced-fault controls across 200, 220, and 300 ms recovered within 23--42 ms despite deeper observed late-fault sags than the corresponding SCR-2 baseline cases. These results challenge a fault-depth-only interpretation but do not isolate an internal synchronization mechanism or establish a universal clearing-time threshold. Reporting the voltage tolerance, measurement-window alignment, clearing instant, qualification interval, and finite observation horizon is essential when comparing EMT recovery trajectories in industrial inverter-based power systems. 

\end{document}